\documentclass[aps,pra,preprint]{revtex4-1}

\usepackage{amsmath}
\usepackage{graphicx}
\usepackage{sistyle}
\usepackage{todonotes}
\usepackage{gensymb}
\usepackage{color}
\usepackage{bm}
\usepackage{dsfont}
\usepackage{braket}
\usepackage{epsfig}
\usepackage{textcomp}
\usepackage{mathtools}
\usepackage{natbib}
\begin{document}
\title{Properties of upconverted entangled broadband light}
\date{\today}
\author{Stefan Lerch}
\author{Andr\'{e} Stefanov}
\email{andre.stefanov@iap.unibe.ch}

\affiliation{University of Bern, Institute of Applied Physics, Sidlerstrasse 5, 3012 Bern, Switzerland}

\keywords{Nonlinear optics, parametric processes;Quantum optics, Photon statistics}

\begin{abstract} 
We experimentally investigate the properties of light generated by the sequential processes of spontaneous parametric downconversion followed by upconversion. By interfering the upconverted photons with the original pump light, we  show that the processes preserve the coherence of the light. Moreover, we observe interferences with doubled phase sensitivity when changing the phase of the downconverted entangled light, although its coherence length is several orders of magnitude shorter than the one of the pump light. Finally the statistics of the upconverted light is expected to be super-poissonian and we experimentally observe superbunching.
\end{abstract}

\maketitle

\section{Introduction}

Entanglement is a resource allowing to potentially overcome the limitations of classical physics in many applications \cite{OBrien2010}. In particular, quantum metrology makes use of specific quantum features to improve the precision of measurements \cite{Giovannetti2011a}. For the estimation of energy-time properties, like phase, time delays or spectroscopic characteristics \cite{Dorfman2016a}, photons entangled in energy and time \cite{Franson1989} are the system of choice. When produced pairwise by spontaneous parametric down-conversion (SPDC) \cite{Klyshko1988} they show the particular feature to exhibit simultaneously narrowband and broadband attributes \cite{Dayan2004}. Indeed, whereas  the energies of the two photons of the pair sum up to the energy of the pump photon, which can originates from a monochromatic laser, the individual spectrum of each photon can be very broad, depending on the specific phase matching conditions in the SPDC crystal. Beside entangled states, the other class of states of light relevant for metrology are states with statistical and coherence properties different from coherent or thermal light. For instance states exhibiting superbunching can improve imaging resolution \cite{Hong2012} or increase the two-photon excited fluorescence rate for microscopy \cite{Jechow2013}. As an example, bright squeezed vacuum with a high mean number of photon per mode shows superbunching \cite{Sh.Iskhakov2012}.

We investigate here the regime of low gain broadband emission SPDC. Spectral width of a few tens of nanometers can be achieved routinely by SPDC, and even broader spectra can be generated by means of dedicated phase-matching \cite{Nasr2008}. The corresponding coherence time of such entangled photons is thus in the femtosecond range. The usual electronic coincidence detection with picosecond or nanosecond resolution cannot be used to capture the short time characteristics of such states of light, and has to be replaced by an optical ultrafast detection. This is realized by sum-frequency generation (SFG) of the two photons in the same crystal (up-conversion) \cite{Dayan2005, ODonnell2009, Zah2008}, or by SFG of each photon independently with a short optical pulse \cite{Kuzucu2008}.

As an detection method with time resolution shorter than the photons coherence time, SFG is sensitive to the relative phase between the two photons \cite{ODonnell2011}. Non-local dispersion cancellation can also be observed for even or for odd  dispersion terms \cite{Lukens2013}. Moreover, the up-converted photons feature further interesting properties. First of all, the up-converted signal can be compared to the original pump light, achieving sensitivity on the global phase of the biphoton. As an example, the SFG process can be stimulated by the pump laser \cite{Shaked2014}. Phase sensitivity of the biphoton is also required for state characterization by homodyne detection as suggested in \cite{Harris2007}. Second, in SFG, the photon statistics of the fundamental waves determine the statistic of the up-converted state \cite{Walls1972}. This is of potential interest because squeezed input states show strong superbunching that can possibly transferred to the narrowband SFG light. Therefore, the successive processes of SPDC followed by up-conversion lead to novel quantum states that are not only of fundamental interest, but can possibly find applications in imaging or two-photon fluorescence.

In this work we investigate the properties of the light resulting from the upconvertion of broadband energy-time entangled photon states. At first, we verify the temporal coherence of the process by letting the resulting photons interfere with the original pump light. We probe the phase experienced by the down-converted biphoton and observe interferences with a doubled sensitivity compared to light at the same wavelength, similarly to what is achieved by interferometery with N00N states \cite{Dowling2008}. Remarkably, those interferences are present although the interferometer path difference is five orders of magnitude larger than the coherence length of the down-converted light. This is a consequence of the dual aspect of energy-time entangled photons possessing a broadband single photon spectrum but long biphoton coherence time. Furthermore, we investigate the statistical properties of the up-converted state of light by measuring its second-order correlation with a Hanbury Brown-Twiss (HBT) interferometer \cite{Brown1957}. A degree of second-order coherence of $8.5\pm 1.4$ is observed, showing strong superbunching.

\section{Theory}
\label{sec:theoryUCpaper}

\begin{figure}[!ht]
\centering
\includegraphics[width=0.6\columnwidth]{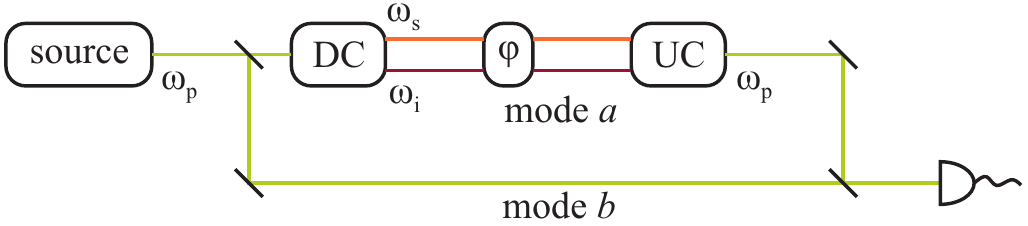}
\caption{A simplified block diagram of the experiment, showing the interferometric measurement. The pump photon is sent into an interferometer. In mode $a$, the photon generates a biphoton by SPDC. The biphoton experiences a phase retardation and gets annihilated by up-conversion (UC). In mode $b$, the pump photon propagates unaffected. The up-converted photon in mode $a$ and the pump photon in mode $b$ are finally recombined and detected.}
\label{fig:schematics}
\end{figure}

At first, we describe the propagation of a monochromatic pump photon through a Mach-Zehnder-like interferometer including the down- and up-conversion processes (Fig. \ref{fig:schematics}). We then discuss the specific photon number distribution of the different states of light in the interferometer.

\subsection{Single Photon Interferences}
We assume first that the pump and the up-converted light are described by a single photon state. It is justified by the low mode population at the detection (well bellow one). In addition, we consider an infinitely narrow pump spectrum of frequency $\omega_p$. After the first beam splitter, the state in mode $a$ and $b$ reads
\begin{equation}
	\ket{\Psi}=\frac{\ket{1}^a_{\omega_p}+\ket{1}^b_{\omega_p}}{\sqrt{2}}
\end{equation}
with $\ket{1}^{a,b}_{\omega_p}= \hat{a}^\dag_{a,b}(\omega_p)\ket{0}$.
In mode $a$, the single photon state gets then annihilated and signal and idler photons are created by SPDC in a state described by a joint spectral amplitude (JSA) $\Lambda(\omega_i,\omega_s)$. In general the JSA is not factorizable. Its exact form depends on the phase-matching condition in the nonlinear crystal. In the monochromatic pump approximation and because of energy conservation ($\omega_p=\omega_i+\omega_s$), we write $\omega_i=\omega_0+\Omega$, $\omega_s=\omega_0-\Omega$ with $\omega_0=\omega_p/2$, and denote the JSA dependency as  $\Lambda(\omega_0,\Omega)$. The state in arm $a$ after down-conversion and blocking of the pump photons is thus
\begin{equation}
\ket{2}^a_{is}=\int\mathrm{d}\Omega\;\Lambda(\omega_0,\Omega)\ket{1}^a_{\omega_0+\Omega}\ket{1}^a_{\omega_0-\Omega},
\end{equation}
which is a two-photon entangled state. At this point, the global state in the interferometer can be written as
\begin{equation}
\ket{2}^a_{is}\ket{0}_{\omega_p}^b+\ket{0}^a_{is}\ket{1}_{\omega_p}^b,
\label{eq:NOOI}
\end{equation}
which describes a superposition of two photons in mode $a$ together with zero photons in mode $b$, and zero photons in mode $a$ together with one photon in mode $b$. Recalling the N00N state notation, where $N$ particles are in a superposition of being all in one mode $a$ and zero particles in mode $b$ or vice versa, this state could be called a N00I state with $N=2$. The biphoton state experiences then a transformation described by the transfer function $H(\omega_i,\omega_s)=H(\omega_0,\Omega)$
\begin{equation}
\ket{2}_{is}^a\rightarrow\int\mathrm{d}\Omega\;\Lambda(\omega_0,\Omega)H(\omega_0,\Omega)\ket{1}^a_{\omega_0+\Omega}\ket{1}^a_{\omega_0-\Omega}.
\end{equation}
If the biphoton propagates through a linear medium the transfer function is factorizable  $H(\omega_i,\omega_s)=H_i(\omega_i) H_s(\omega_s)$, but it can be a non-factorizable function when nonlinearities are involved, as in the case of two-photon spectroscopy. In our experiment, the transfer function is simplified by only considering a phase shift $\varphi(\omega_{i,s})$ such that the state after phase shifting is
\begin{equation}
\int\mathrm{d}\Omega\;\Lambda(\omega_0,\Omega)e^{i[\varphi(\omega_o+\Omega)+\varphi(\omega_0-\Omega)]}\ket{1}^a_{\omega_0+\Omega}\ket{1}^a_{\omega_0-\Omega}.
\end{equation}

In the next step, signal and idler are up-converted by the process of SFG into a single photon with frequency $\omega_p$ whose state is given by $g(\omega_p)\ket{1}_{\omega_p}^a$ with the complex amplitude factor
\begin{equation}
g(\omega_p)=\int\mathrm{d}\Omega\;\Gamma(\omega_0,\Omega)e^{i[\varphi(\omega_0+\Omega)+\varphi(\omega_0-\Omega)]}.
\label{eq:g_state}
\end{equation}
The function $\Gamma(\omega_0,\Omega)$ absorbs the phase-matching condition of the up-conversion crystal and the JSA. The state (\ref{eq:g_state}) is not normalized in order to describe changes in both amplitude and phase of the biphoton. However, the overall efficiencies of the SPDC and SFG process are not taken into account in this description. The generated photon is indistinguishable from the pump photon, and therefore, when finally the two paths are recombined, the state at the output of the interferometer reads 
\begin{equation}
\frac{1+g(\omega_p)}{2}\ket{1}_{\omega_p}^a.
\end{equation}
The detection probability at one output of the interferometer is proportional to
\begin{equation}
p \propto\left|1+g(\omega_p)\right|^2.
\end{equation}
This probability is obviously sensitive to the dispersion relation $\varphi(\omega)$. By expanding the phase around $\omega_0$
\begin{equation}
\varphi(\omega_0\pm\Omega)=\varphi(\omega_0)\pm\alpha\Omega+\frac{\beta}{2}\Omega^2\pm\frac{\gamma}{6}\Omega^3+\mathcal{O}(\Omega^4),
\end{equation}
we see that $g(\omega_p)$ is insensitive to the odd orders of the expansion and has twice the sensitivity for the even orders. Assuming that all second and higher orders of the dispersion are compensated, the linear term in the phase corresponds to the phase which would be induced by a length change $z$ of the interferometer arm $\varphi(\omega)=\omega z/c$. As a consequence, we find $g(\omega_p)\propto\mathrm{exp}\{2i\omega_0 z/c\}$ such that the interference fringes are given by
\begin{equation}
p \propto1+\cos\left(\frac{2\omega_0 z}{c}\right).
\label{eq:pump_sensitivity}
\end{equation}
They have twice the periodicity of the down-converted photon central wavelength. Moreover, the probability is independent on the bandwidth of the down-converted photons. 

\subsection{Photon Statistics}\label{sec:photon_statistics_ucpaper}
Although both processes, SPDC and up-conversion, are unitary, they are not the unitary conjugate of each other when the pump beam is blocked between the processes. Therefore, the state of light resulting from SFG is not the same as the one from the original pump. The difference can be observed in the photon statistics.

The rather low efficiency of the SPDC and SFG process on the order of $10^{-7}$ makes the experiment with a single photon source virtually impossible. Experimentally, the SPDC process has to be pumped strongly by a laser to get reasonable signal after the sequential nonlinear interactions. As a consequence, the pump source can be considered as a classical field. In mode $b$, the source is attenuated to a small mean photon number $\overline{n}=|\alpha|^2<1$, and can be considered as single mode coherent state \cite{Glauber1963}
\begin{equation}
\ket{\alpha}=\mathrm{e}^{-|\alpha|^2/2}\sum\limits_{n=0}^\infty \frac{\alpha^n}{\sqrt{n!}}\ket{n}
\end{equation}
with Poissonian photon number distribution. In mode $a$, the state after SPDC is given by applying the squeezing operator $\hat{S}(\zeta)=\mathrm{exp}\{[\zeta^*\hat{a}_\mathrm{SPDC}^2-\zeta\hat{a}_\mathrm{SPDC}^{\dag 2}]/2\}$ on the vacuum:
\begin{equation}
\ket{\Psi_{\mathrm{SPDC}}}=\left[\sum\limits_{n=0}^\infty c_n\ket{2n}_\mathrm{SPDC}\right]\ket{0}_\mathrm{SFG},
\end{equation}
\begin{equation}
c_n=\frac{\sqrt{\mathrm{sech}(|\zeta|)(2n)!}}{n!}\left(-\frac{\mathrm{e}^{i\phi}}{2}\mathrm{tanh}(|\zeta|)\right)^n,
\end{equation}
where $\zeta=|\zeta|\mathrm{e}^{i\phi}$ quantifies the strength of the nonlinearity and of the pump intensity \cite{Loudon1987}. We only consider two modes of light for simplicity, the one of the photon pairs and the one of the up-converted photons. The SPDC, being broadband, is highly multimode in our experiment. However the phase-matching condition for long nonlinear crystals makes it highly improbable to observe SFG from SPDC photons of different modes. This validates the two mode treatment of the sequential process. The SFG is then described by the operator $\hat{S}_\mathrm{SFG}(\kappa)=\mathrm{exp}\{\kappa(\hat{a}_\mathrm{SPDC}^2\hat{a}_\mathrm{SFG}^\dag-\hat{a}_\mathrm{SPDC}^{\dag 2}\hat{a}_\mathrm{SFG})\}$ acting on $\ket{\Psi}_\mathrm{SPDC}$:
\begin{equation}
\ket{\Psi_\mathrm{SFG}}=\hat{S}_\mathrm{SFG}(\kappa)\ket{\Psi_{\mathrm{SPDC}}}.
\end{equation} 
Here, $\kappa$ quantifies the nonlinear interaction strength. The state of the SFG mode is then obtained by tracing out the SPDC mode
\begin{equation}
\hat{\rho}_\mathrm{SFG}=\mathrm{Tr}_\mathrm{SPDC}\{\ket{\Psi_\mathrm{SFG}}\bra{\Psi_\mathrm{SFG}}\}.
\end{equation} 
\begin{figure}[tbh]
\centering
\includegraphics[width=0.3\columnwidth]{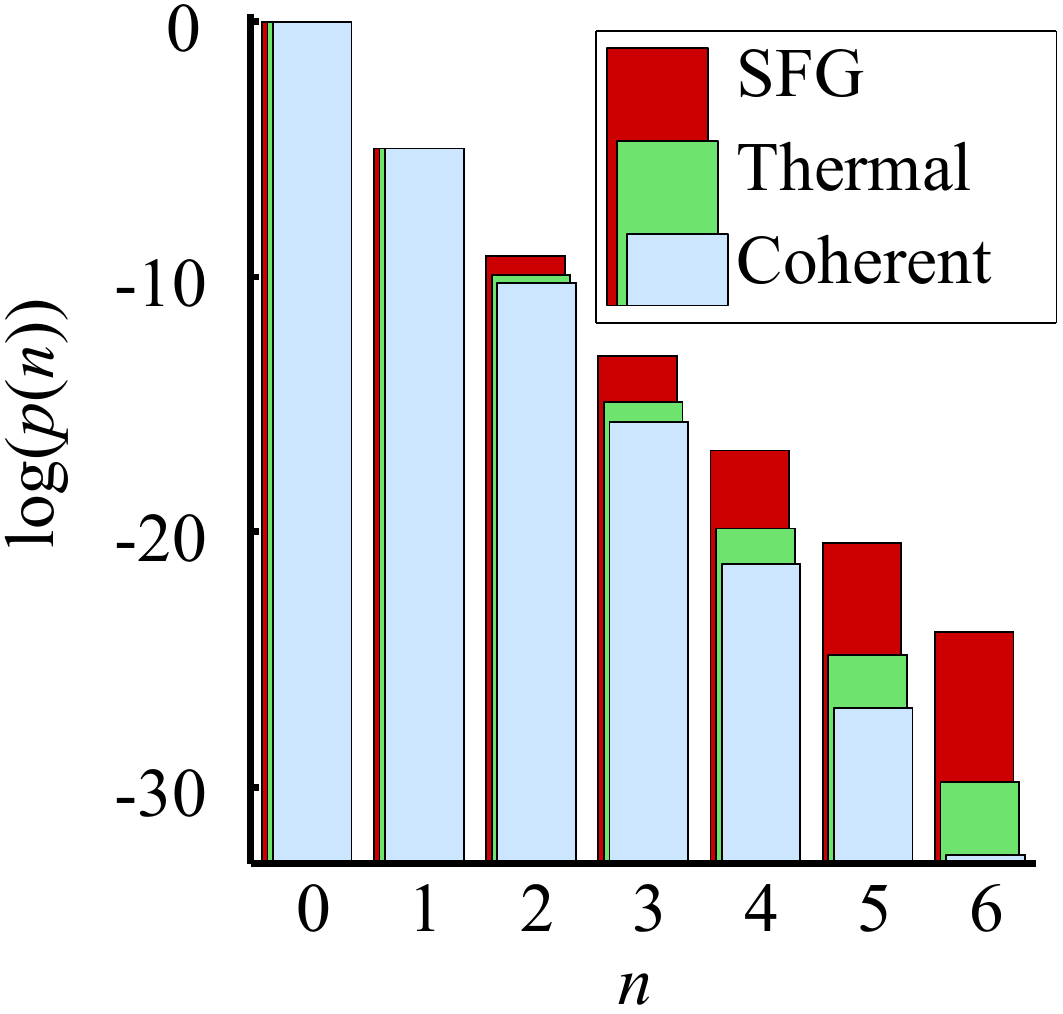}
\caption{Photon number distribution of the state of light after SFG (red, $g^{(2)}_\mathrm{SFG}=11.3$). The photon number distribution of a thermal state (green, $g^{(2)}_\mathrm{th}=2.0$) and a coherent state (blue, $g^{(2)}_\mathrm{coh}=1.0$) with equal mean photon number are shown for comparison.}
\label{fig:p_n_theory}
\end{figure}
For small photon numbers, the state can be numerically computed using a truncated Fock basis for each mode. Exemplary, we calculate the SFG state up to the first 50 Fock states with parameters that reproduce the intensities observed in the experiment, i.e. $\overline{n}_\mathrm{SPDC}=\mathrm{sinh}(|\zeta|)^2=0.1$, and $\overline{n}_\mathrm{SFG}=10^{-5}$ which imposes $\kappa=8.7715\times 10^{-3}$. Figure \ref{fig:p_n_theory} shows the photon number distribution $p(n)=\prescript{}{\mathrm{SFG}}{\braket{n|\hat{\rho}_\mathrm{SFG}|n}_\mathrm{SFG}}$ up to $n=6$ for the up-converted mode after SFG, and, for comparison, for coherent and thermal states with the same mean photon number. According to this simple model, the SFG state is expected to show superbunching with a value of the second-order correlation as high as $g_{\mathrm{SFG}}^{(2)}=11.3$, which is much larger than for thermal states where $g_{\mathrm{th}}^{(2)}=2$. The numerical simulation is in agreement with the expected result for low mean photon numbers, i.e. the second-order coherence of the photons created in a two-photon-annihilation process is equal to the ratio of the fourth-order coherence of the incoming photon to the second-order coherence squared:
\begin{equation}
g^{(2)}_\mathrm{SFG}=\frac{g^{(4)}_\mathrm{SPDC}}{\left(g^{(2)}_\mathrm{SPDC}\right)^2}.
\end{equation}
The calculation reveals that the action of $\hat{S}_\mathrm{SFG}(\kappa)$ slightly entangles the SPDC and the SFG mode, i.e. $\hat{\rho}_\mathrm{SFG}$ is not pure, but $\mathrm{Tr}\{\hat{\rho}_\mathrm{SFG}^2\}=0.999997$. Finally, the fidelity to a coherent state with equal mean photon number is $F=\mathrm{Tr}\{\sqrt{\sqrt{\hat{\rho}_\mathrm{SFG}}\ket{\alpha}\bra{\alpha}\sqrt{\hat{\rho}_\mathrm{SFG}}}\}=99.998$\%, i.e. mode $a$ and mode $b$ are almost indistinguishable and can interfere.
%of $\hat{\rho}=(\hat{\rho}_\mathrm{SFG}+\ket{\alpha}\bra{\alpha})/2$ with $\overline{n}_\mathrm{SFG}=|\alpha|^2$ yields a purity of $\mathrm{Tr}\{\hat{\rho}^2\}=99.998$\%.

\section{Experiments}\label{sec:experimentUCpaper}

\subsection{Interferences Between Pump and SFG Light}\label{sec:setupUCpaper}
\begin{figure}[t]
\centering
\includegraphics[width=0.7\columnwidth]{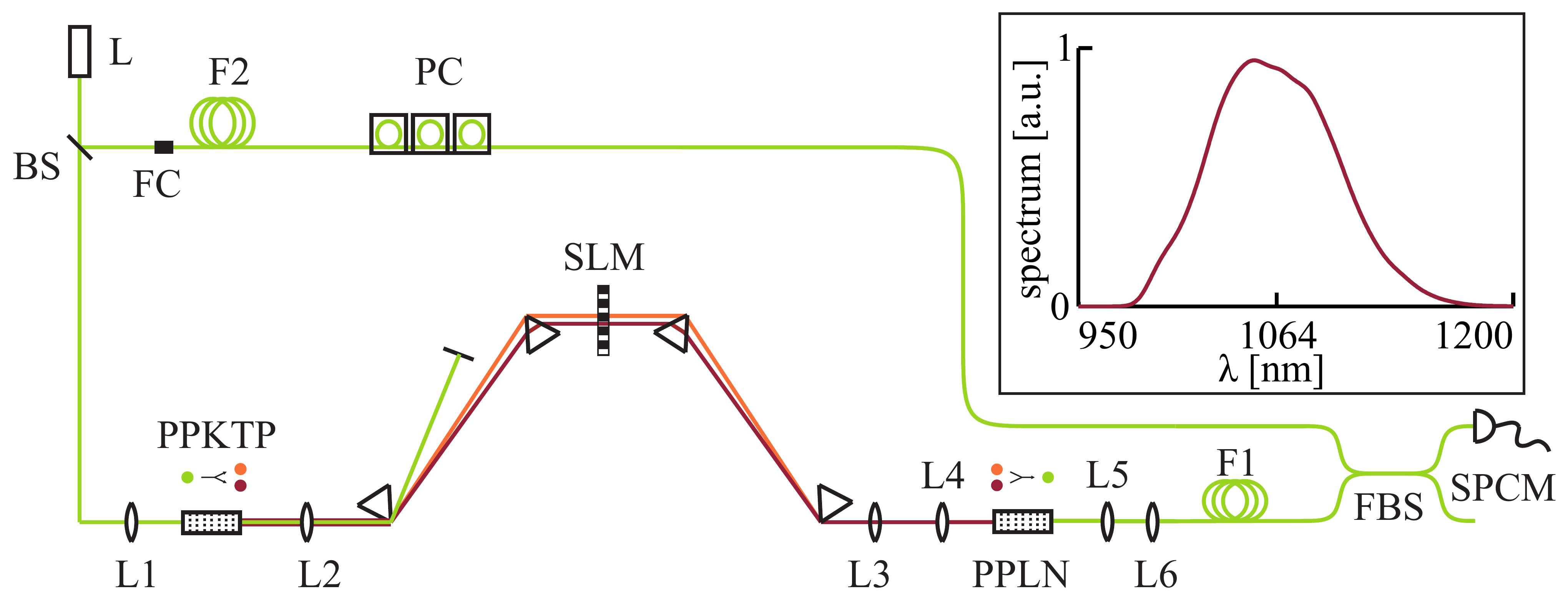}
\caption{Schematic of the experimental setup. The pump laser (L) is splitted by a beam sampler (BS). Most of the power is transmitted and focused by a lens (L1) into the PPKTP crystal and pumps the SPDC process. The generated pairs are imaged by a lens to the symmetry plane of a four-prism compressor, where an SLM is placed. This plane is imaged by lenses L3 and L4 into the PPLN crystal. The up-converted photons are coupled with the lenses L5 and L6 into a single mode fiber (F1). The reflected part of the BS is coupled with a fiber coupler (FC) into a single mode fiber (F2). The polarization is controlled by means of a polarization controller (PC). Both arms are recombined in a fiber coupler (FBS) and sent to the fiber coupled single photon counting module (SPCM). The inset shows the down-conversion spectrum measured with a 200 \textmu m fiber at the position of the single mode fiber F1. The spectrometer resolution is 10~nm.}
\label{fig:setup}
\end{figure}

We experimentally implement the Mach-Zehnder-like interferometer discussed in the previous section. Its schematic is shown in Fig. \ref{fig:setup}. %Observing such interferences is the first step toward complete characterization of the function $g(\omega_p)$ by homodyne detection and it demonstrates that the coherence is preserved between the processes of SPDC and SFG.%
A monochromatic pump laser (Verdi, 4~W power) at 532~nm is split into the two arms of the interferometer by means of a beam sampler. One arm (mode $b$) is coupled into a single mode fiber and sent through a Lefevre three loop polarization controller \cite{Lefevre1980}, before it is recombined with the second arm at a 50/50 fiber beam splitter. The light in the second arm (mode $a$) pumps a 10~mm long periodically poled KTP (PPKTP) crystal, where the phase-matching is chosen for mainly collinear type-0 SPDC \cite{Lerch2014}. The generated biphotons are degenerate and broadband around 1064~nm (770~nW power). Their spectrum is depicted as an inset in Fig. \ref{fig:setup}. A four-prism compressor allows to compensate for dispersion. The spatially dispersed frequency spectrum can be shaped by means of a spatial light modulator (SLM) \cite{Zah2008}. In this experiment we use the SLM purely as a variable phase retarder, i.e. it allows to introduce an arbitrary phase $\varphi$ on the biphoton state. We chose it to be constant for all wavelength. After passing the prism compressor, the biphoton undergoes up-conversion in a periodically poled lithium niobate (PPLN) crystal. The phase-matching of the PPLN and the PPKTP differs only marginally, but the PPLN provides a higher interaction strength. The up-converted photons are coupled into a single mode fiber, and combined with the interferometer arm $b$ by a single mode fiber coupler to guarantee spatial overlap. Finally, one of the outputs of the interferometer is detected by a single photon counting module (Perkin Elmer SPCM-AQR-15). The coupling of light into arm $b$ is set such that the detected mean photon rate with blocked arm $a$ is similar to the detected up-conversion rate. The raw mean rate in arm $a$, where up-conversion takes place, is 344.4~Hz while it is 568.8~Hz in the other arm. The dark count rate is 202.9~Hz. The lengths of the interferometer arms are highly unbalanced. The photons in the upper arm (mode $b$) propagate mainly in about 7~m single mode fiber, and in the lower arm (mode $a$), they propagate through approximately 3~m in free space but only one meter in fiber. The corresponding path length difference in air is 1~m. The interferometer is subjected to various random phase drifts. In particular the ovens, controlling the temperatures of the two bulk crystals, produce strong air convection, and thus induce random dephazing. Because the interferometer is not stabilized, the phase scan has to be performed faster than the time scale in which the random phase change happens such that the visibility and the phase sensitivity of the interference fringes can be measured. We scan the phase by steps of $\pi/9$ with an acquisition time of 100 ms for each step. Another 100 ms between the individual steps are required to change the phase setting by the SLM. 

\begin{figure}[t]
\centering
\includegraphics[width=.5\columnwidth]{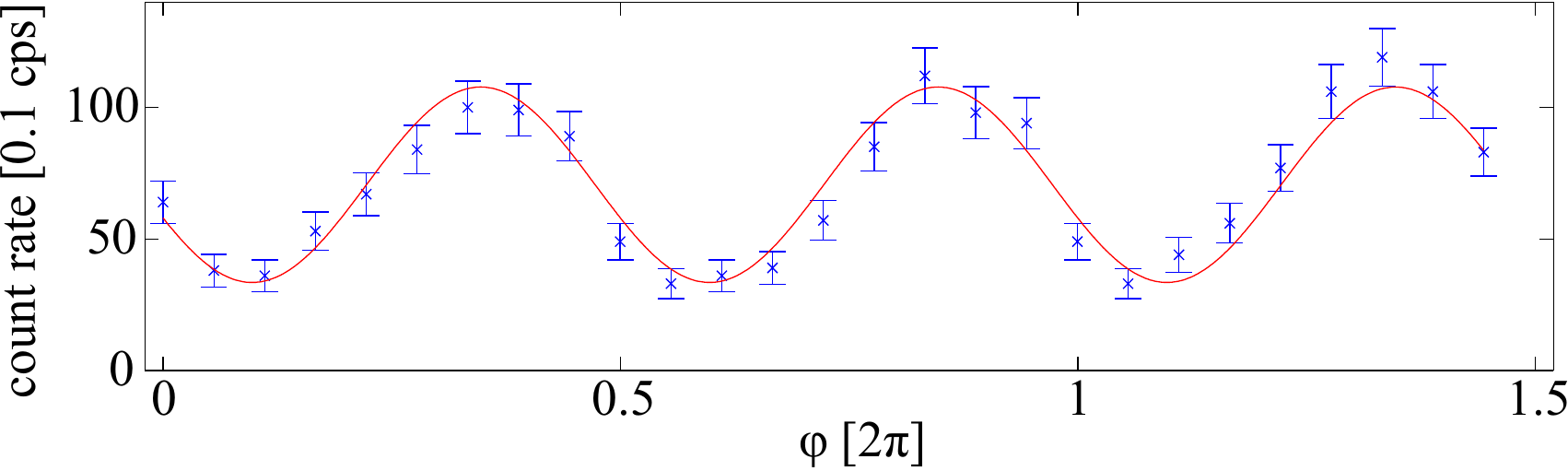}
\caption{Interference between up-conversion photons and pump photons when the phase retardation $\varphi$ in mode $a$ (between SPDC and up-conversion) is changed. The raw measurement data (blue) and the fit with fixed frequency of 0.5 periods${}^{-1}$ (red solid curve) illustrate the doubled phase sensitivity. The errorbars are calculated assuming Poissonian distributed count rates.}
\label{fig:result}
\end{figure}

We observe interference between up-conversion and pump photons when the phase $\varphi$, experienced by the down-converted photons, is varied over $3\pi$. The raw count rates are shown in Fig. \ref{fig:result}. Subtracting the dark counts, we obtain a visibility of $V=(73.8 \pm 5.1)$\%. The maximal achievable visibility is however only $V_\mathrm{max}=89.7$\% because of the unbalanced intensities between the two arms. The discrepancy means that the pump photon and the SFG photons are distinguishable. The numerical calculation preclude a visibility reduction because of different photon statistics. The visibility can be however limited by several other factors. First, because of the low count rates, it is difficult to optimally adjust the polarization overlap at the last beam splitter. Second, the coherence length of the pump laser, although large, could still limit the visibility when the optical path length difference is about 1 m as in the experiment. The pump laser is specified to have a bandwidth smaller than 5 MHz at 50 ms integration time. The corresponding coherence length is about 17 m. However we integrate over longer times, such that the effective coherence length could be shorter. The indistinguishability of the pump photons and the photons after SPDC and SFG is therefore probably larger than $V/V_\mathrm{max}=(82.3\pm5.7)$\%.

Signal and idler exhibit a broad spectrum. The corresponding short coherence length is about 15 \textmu m (50 fs coherence time). Thus, an interesting feature of the interferometer is the interim incoherent nature of the light in one arm of the interferometer, while exhibiting high visibility with large unbalancing. This property is a signature of the strong energy correlations between the photons of the pairs.

Beside the high visibility, the most striking feature, illustrated by Fig. \ref{fig:result}, is the doubled sensitivity to the phase retardation $\varphi$ introduced by the SLM. Similarly to a N00N interferometer,  the phase sensitivity is here multiplied by $N$, the number of photons which probe simultaneously the phase. Whereas the generation of N00N states requires an element that splits $N$ incoming particles between the two modes, here, the nonlinear crystal is placed within the interferometer, producing a N00I state as shown by Eq. (\ref{eq:NOOI}). While the proof of concept is here realized for $N=2$ by making use of second-order nonlinearities, going to higher order nonlinear processes would allow to increase the sensitivity even further. In this example, the phase change accumulated by one photon is constant for all wavelength. Therefore, the sensitivity is twice as much as for an interferometer based on pump photons. On the contrary, a phase variation induced by a length change between down- and up-conversion crystal would lead to the same phase sensitivity as for interferometry with the pump. Equation \eqref{eq:pump_sensitivity} illustrates this, as we identify $2\omega_0=\omega_p$. The intrinsically low interaction strengths, the low count rates, respectively, make this scheme of rather difficult use. However, our interferometer has potential when partially opaque media are studied. For a medium that is opaque for the pump mode but transparent for the down-converted modes, phase changes can still be observed with the absolute resolution that is determined by the pump.

%\section{Results and Discussion}

\subsection{Second-Order Coherence Measurements}\label{sec:G2}
\begin{figure}[t]
\centering
\includegraphics[width=.7\columnwidth]{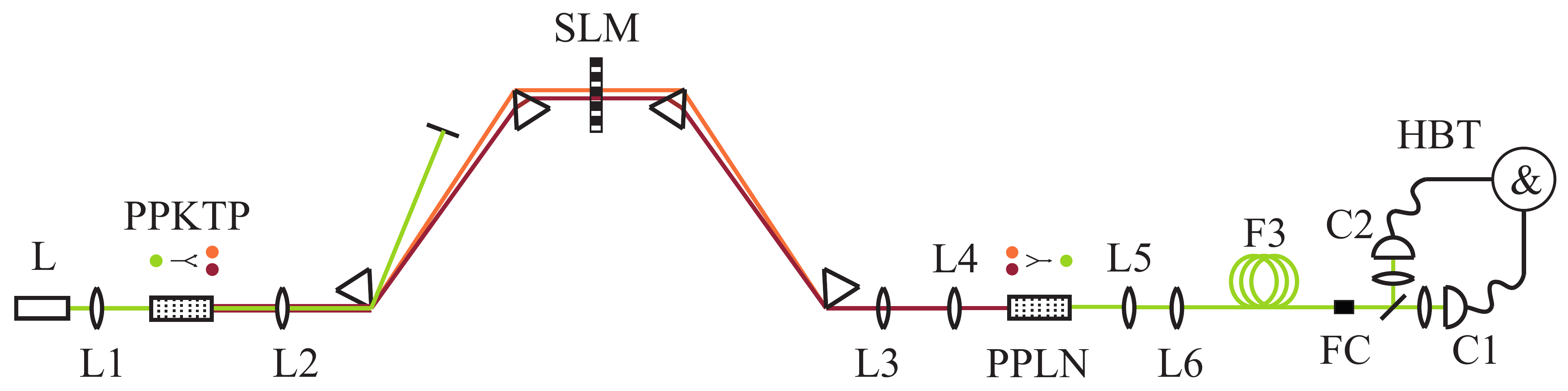}
\caption{Schematic of the second-order coherence measurement. Similar to the setup in Fig. \ref{fig:setup}, SPDC and SFG happen sequentially. The SFG light is coupled with the lenses L5 and L6 into a fiber (F3). A fiber coupler FC sends the photons into a free space HBT interferometer. The interferometer consists of a 50/50 beam splitter, two lenses, and two single photon detectors (C1 and C2, ID Quantique id100-50). The detector signals are recorded and checked for coincidence counts.}
\label{fig:HBT_setup}
\end{figure}

As shown of Fig. \ref{fig:p_n_theory}, while the photon number distribution of the pump state is Poissonian, super-Poissonian statistics after SFG is expected. In order to test it, we implement a Hanbury Brown-Twiss (HBT) interferometer after the up-conversion crystal. Its schematic is depicted in Fig. \ref{fig:HBT_setup}. Similar to the Mach-Zehnder-like interferometer, we pump the PPKTP crystal with the Verdi laser. The down-converted light propagates through the prism compressor setup, and enters the second crystal where it undergoes up-conversion. Two lenses couple the up-converted light into a fiber, which guides it to a free space HBT interferometer. A fiber coupler collimates the up-converted light. It is separated by a 50/50 beam splitter, and focused by lenses (focal length 11 mm and 12 mm) onto two single photon detectors (ID Quantique id100-50). The two individual detectors measure a raw mean rate of 60.9~Hz (50.3~Hz) with a dark count rate of 8.0~Hz (5.8~Hz). The dead time amounts 45~ns. The signals are measured during 40~hours leading to $1.6\times 10^7$ detection events. The high dark count rate of the fiber coupled detector enforces the use of a free space alternative. Even though we lose signal at each element in the HBT interferometer, the signal-to-noise ratio is considerably better. In order to increase the ratio further, we use a few mode fiber (single mode fiber for 633 - 780~nm) for guiding the up-converted light into the HBT interferometer instead of a single mode fiber. This increases the coupling efficiency, respectively the count rate, but still guarantees a good spatial overlap. The acquisition electronic (PCIe-6321 from National Instruments) assigns a time stamp to each individual event with a resolution of 10 ns. In order to obtain the second-order coherence, we calculate for each detection event of one detector the time differences to the detection events of the other detector. The frequency of events with time difference $\tau$ is then proportional to $g^{(2)}(\tau)$. 

\begin{figure}[t]
\centering
\includegraphics[width=.7\columnwidth]{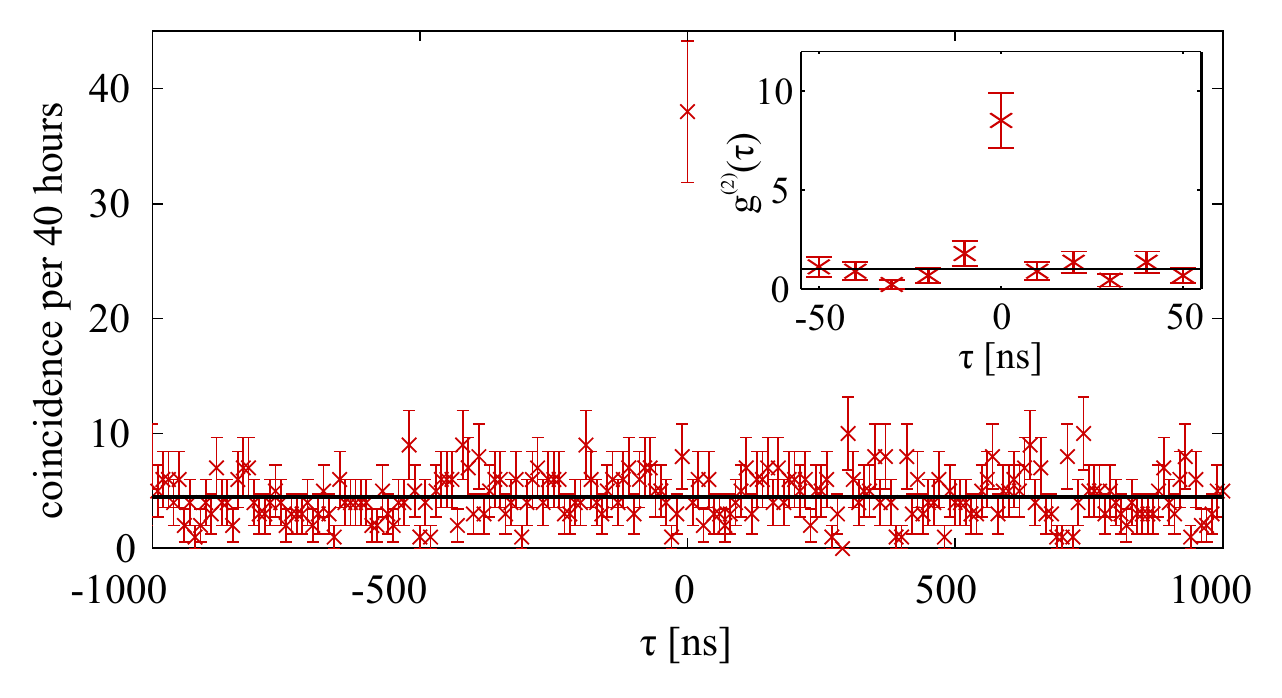}
\caption{Absolute coincidence counts within 40 hours (second-order correlation function) as a function of the time difference between two detection events at counter C1 and C2 (see Fig. \ref{fig:setup}). The peak of ($38\pm 6$) counts at $\tau=0$ ns exceeds clearly the average background of ($4.46\pm 0.14)$ counts (black solid line). The errors are calculated assuming Poissonian statistics. The inset zooms in the region of $\tau\in[-55,55]$ ns. The coincidence events are rescaled to $g^{(2)}(\tau)$ assuming the background to be one.}
\label{fig:G2}
\end{figure}

The result of the HBT measurement is shown in Fig. \ref{fig:G2}. A strong peak at zero time delay is observed with $g^{(2)}(0)=8.5\pm 1.4$ demonstrating superbunching of light. Apparently, the temporal resolution of 10 ns is too low to resolve the shape of the peak, but still good enough not to wash it out. This indicates that the second order coherence length of the upconverted light is of the order of a few nanoseconds. The SFG light features an inherently narrow bandwidth because of the narrow pump beam. However, linking the width of $g^{(2)}(\tau)$ to the pump coherence would requires a multimode evaluation of both SPDC and SFG. A numerical treatment of the corresponding multimode process, similar to the one presented in this paper, is impossible because the number of possible down-conversion modes is on the order of $10^6$ \cite{Bernhard2013}. However, if the width of the second-order coherence of the SFG photons is directly related to the pump bandwidth, we anticipate this second order correlation measurement as a possible measure for the coherence time of narrowband lasers.

\section{Conclusion}
\label{sec:conclusionUCpaper}

Our results demonstrate the coherence of the up-converted light from SPDC with the pump laser. As a consequence it is possible to observe interferences with high visibility even though the light in the interferometer is of very short coherence. In addition, the sensitivity to a phase shift, experienced by the down-converted light, is doubled with respect to classical light. Finally measuring the second-order correlation function of the up-converted light demonstrates strong superbunching. %While only a complete tomographical reconstruction \cite{Breitenbach1997} would fully characterize the state, the fact that up-conversion is shown to be a coherent process together with the expected superbunching provides confidence that the up-converted state is a coherent superposition of Fock states with super-poissionian photon number distribution. 
It is an open question how the width of the second-order correlation is linked to the original pump bandwidth. Such new kind of light can find promising applications in phase estimation, super-resolution imaging and two-photon fluorescence.

\section*{Funding Information}
This research was supported by the Swiss National Science Foundation (Grant No. No.~PP00P2\_159259).

\bibliographystyle{apsrev4-1} 

\begin{thebibliography}{26}%
\makeatletter
\providecommand \@ifxundefined [1]{%
 \@ifx{#1\undefined}
}%
\providecommand \@ifnum [1]{%
 \ifnum #1\expandafter \@firstoftwo
 \else \expandafter \@secondoftwo
 \fi
}%
\providecommand \@ifx [1]{%
 \ifx #1\expandafter \@firstoftwo
 \else \expandafter \@secondoftwo
 \fi
}%
\providecommand \natexlab [1]{#1}%
\providecommand \enquote  [1]{``#1''}%
\providecommand \bibnamefont  [1]{#1}%
\providecommand \bibfnamefont [1]{#1}%
\providecommand \citenamefont [1]{#1}%
\providecommand \href@noop [0]{\@secondoftwo}%
\providecommand \href [0]{\begingroup \@sanitize@url \@href}%
\providecommand \@href[1]{\@@startlink{#1}\@@href}%
\providecommand \@@href[1]{\endgroup#1\@@endlink}%
\providecommand \@sanitize@url [0]{\catcode `\\12\catcode `\$12\catcode
  `\&12\catcode `\#12\catcode `\^12\catcode `\_12\catcode `\%12\relax}%
\providecommand \@@startlink[1]{}%
\providecommand \@@endlink[0]{}%
\providecommand \url  [0]{\begingroup\@sanitize@url \@url }%
\providecommand \@url [1]{\endgroup\@href {#1}{\urlprefix }}%
\providecommand \urlprefix  [0]{URL }%
\providecommand \Eprint [0]{\href }%
\providecommand \doibase [0]{http://dx.doi.org/}%
\providecommand \selectlanguage [0]{\@gobble}%
\providecommand \bibinfo  [0]{\@secondoftwo}%
\providecommand \bibfield  [0]{\@secondoftwo}%
\providecommand \translation [1]{[#1]}%
\providecommand \BibitemOpen [0]{}%
\providecommand \bibitemStop [0]{}%
\providecommand \bibitemNoStop [0]{.\EOS\space}%
\providecommand \EOS [0]{\spacefactor3000\relax}%
\providecommand \BibitemShut  [1]{\csname bibitem#1\endcsname}%
\let\auto@bib@innerbib\@empty
%</preamble>
\bibitem [{\citenamefont {O'Brien}\ \emph {et~al.}(2010)\citenamefont
  {O'Brien}, \citenamefont {Furusawa},\ and\ \citenamefont
  {Vu{\v{c}}kovi{\'{c}}}}]{OBrien2010}%
  \BibitemOpen
  \bibfield  {author} {\bibinfo {author} {\bibfnamefont {J.~L.}\ \bibnamefont
  {O'Brien}}, \bibinfo {author} {\bibfnamefont {A.}~\bibnamefont {Furusawa}}, \
  and\ \bibinfo {author} {\bibfnamefont {J.}~\bibnamefont
  {Vu{\v{c}}kovi{\'{c}}}},\ }\href {\doibase 10.1038/nphoton.2009.229}
  {\bibfield  {journal} {\bibinfo  {journal} {Nature Photonics}\ }\textbf
  {\bibinfo {volume} {3}},\ \bibinfo {pages} {687} (\bibinfo {year}
  {2010})}\BibitemShut {NoStop}%
\bibitem [{\citenamefont {Giovannetti}\ \emph {et~al.}(2011)\citenamefont
  {Giovannetti}, \citenamefont {Lloyd},\ and\ \citenamefont
  {Maccone}}]{Giovannetti2011a}%
  \BibitemOpen
  \bibfield  {author} {\bibinfo {author} {\bibfnamefont {V.}~\bibnamefont
  {Giovannetti}}, \bibinfo {author} {\bibfnamefont {S.}~\bibnamefont {Lloyd}},
  \ and\ \bibinfo {author} {\bibfnamefont {L.}~\bibnamefont {Maccone}},\ }\href
  {\doibase 10.1038/nphoton.2011.35} {\bibfield  {journal} {\bibinfo  {journal}
  {Nature Photonics}\ }\textbf {\bibinfo {volume} {5}},\ \bibinfo {pages} {222}
  (\bibinfo {year} {2011})}\BibitemShut {NoStop}%
\bibitem [{\citenamefont {Dorfman}\ \emph {et~al.}(2016)\citenamefont
  {Dorfman}, \citenamefont {Schlawin},\ and\ \citenamefont
  {Mukamel}}]{Dorfman2016a}%
  \BibitemOpen
  \bibfield  {author} {\bibinfo {author} {\bibfnamefont {K.~E.}\ \bibnamefont
  {Dorfman}}, \bibinfo {author} {\bibfnamefont {F.}~\bibnamefont {Schlawin}}, \
  and\ \bibinfo {author} {\bibfnamefont {S.}~\bibnamefont {Mukamel}},\ }\href
  {\doibase 10.1103/RevModPhys.88.045008} {\bibfield  {journal} {\bibinfo
  {journal} {Reviews of Modern Physics}\ }\textbf {\bibinfo {volume} {88}},\
  \bibinfo {pages} {045008} (\bibinfo {year} {2016})}\BibitemShut {NoStop}%
\bibitem [{\citenamefont {Franson}(1989)}]{Franson1989}%
  \BibitemOpen
  \bibfield  {author} {\bibinfo {author} {\bibfnamefont {J.}~\bibnamefont
  {Franson}},\ }\href {http://link.aps.org/doi/10.1103/PhysRevLett.62.2205}
  {\bibfield  {journal} {\bibinfo  {journal} {Physical review letters}\
  }\textbf {\bibinfo {volume} {62}},\ \bibinfo {pages} {2205} (\bibinfo {year}
  {1989})}\BibitemShut {NoStop}%
\bibitem [{\citenamefont {Klyshko}(1988)}]{Klyshko1988}%
  \BibitemOpen
  \bibfield  {author} {\bibinfo {author} {\bibfnamefont {D.~N.}\ \bibnamefont
  {Klyshko}},\ }\href {http://books.google.com/books?id=IPfwdhR4TaYC{\&}pgis=1}
  {\emph {\bibinfo {title} {{Photons and nonlinear optics}}}}\ (\bibinfo
  {publisher} {Gordon and Breach},\ \bibinfo {year} {1988})\ p.\ \bibinfo
  {pages} {415}\BibitemShut {NoStop}%
\bibitem [{\citenamefont {Dayan}\ \emph {et~al.}(2004)\citenamefont {Dayan},
  \citenamefont {Pe'er}, \citenamefont {Friesem},\ and\ \citenamefont
  {Silberberg}}]{Dayan2004}%
  \BibitemOpen
  \bibfield  {author} {\bibinfo {author} {\bibfnamefont {B.}~\bibnamefont
  {Dayan}}, \bibinfo {author} {\bibfnamefont {A.}~\bibnamefont {Pe'er}},
  \bibinfo {author} {\bibfnamefont {A.}~\bibnamefont {Friesem}}, \ and\
  \bibinfo {author} {\bibfnamefont {Y.}~\bibnamefont {Silberberg}},\ }\href
  {http://link.aps.org/doi/10.1103/PhysRevLett.93.023005} {\bibfield  {journal}
  {\bibinfo  {journal} {Physical review letters}\ }\textbf {\bibinfo {volume}
  {93}},\ \bibinfo {pages} {23005} (\bibinfo {year} {2004})}\BibitemShut
  {NoStop}%
\bibitem [{\citenamefont {Hong}\ \emph {et~al.}(2012)\citenamefont {Hong},
  \citenamefont {Liu},\ and\ \citenamefont {Zhang}}]{Hong2012}%
  \BibitemOpen
  \bibfield  {author} {\bibinfo {author} {\bibfnamefont {P.}~\bibnamefont
  {Hong}}, \bibinfo {author} {\bibfnamefont {J.}~\bibnamefont {Liu}}, \ and\
  \bibinfo {author} {\bibfnamefont {G.}~\bibnamefont {Zhang}},\ }\href
  {\doibase 10.1103/PhysRevA.86.013807} {\bibfield  {journal} {\bibinfo
  {journal} {Physical Review A}\ }\textbf {\bibinfo {volume} {86}},\ \bibinfo
  {pages} {013807} (\bibinfo {year} {2012})}\BibitemShut {NoStop}%
\bibitem [{\citenamefont {Jechow}\ \emph {et~al.}(2013)\citenamefont {Jechow},
  \citenamefont {Seefeldt}, \citenamefont {Kurzke}, \citenamefont {Heuer},\
  and\ \citenamefont {Menzel}}]{Jechow2013}%
  \BibitemOpen
  \bibfield  {author} {\bibinfo {author} {\bibfnamefont {A.}~\bibnamefont
  {Jechow}}, \bibinfo {author} {\bibfnamefont {M.}~\bibnamefont {Seefeldt}},
  \bibinfo {author} {\bibfnamefont {H.}~\bibnamefont {Kurzke}}, \bibinfo
  {author} {\bibfnamefont {A.}~\bibnamefont {Heuer}}, \ and\ \bibinfo {author}
  {\bibfnamefont {R.}~\bibnamefont {Menzel}},\ }\href {\doibase
  10.1038/nphoton.2013.271} {\bibfield  {journal} {\bibinfo  {journal} {Nature
  Photonics}\ }\textbf {\bibinfo {volume} {7}},\ \bibinfo {pages} {973}
  (\bibinfo {year} {2013})}\BibitemShut {NoStop}%
\bibitem [{\citenamefont {{Sh. Iskhakov}}\ \emph {et~al.}(2012)\citenamefont
  {{Sh. Iskhakov}}, \citenamefont {P{\'{e}}rez}, \citenamefont {{Yu.
  Spasibko}}, \citenamefont {Chekhova},\ and\ \citenamefont
  {Leuchs}}]{Sh.Iskhakov2012}%
  \BibitemOpen
  \bibfield  {author} {\bibinfo {author} {\bibfnamefont {T.}~\bibnamefont {{Sh.
  Iskhakov}}}, \bibinfo {author} {\bibfnamefont {a.~M.}\ \bibnamefont
  {P{\'{e}}rez}}, \bibinfo {author} {\bibfnamefont {K.}~\bibnamefont {{Yu.
  Spasibko}}}, \bibinfo {author} {\bibfnamefont {M.~V.}\ \bibnamefont
  {Chekhova}}, \ and\ \bibinfo {author} {\bibfnamefont {G.}~\bibnamefont
  {Leuchs}},\ }\href {\doibase 10.1364/OL.37.001919} {\bibfield  {journal}
  {\bibinfo  {journal} {Optics Letters}\ }\textbf {\bibinfo {volume} {37}},\
  \bibinfo {pages} {1919} (\bibinfo {year} {2012})}\BibitemShut {NoStop}%
\bibitem [{\citenamefont {Nasr}\ \emph {et~al.}(2008)\citenamefont {Nasr},
  \citenamefont {Carrasco}, \citenamefont {Saleh}, \citenamefont {Sergienko},
  \citenamefont {Teich}, \citenamefont {Torres}, \citenamefont {Torner},
  \citenamefont {Hum},\ and\ \citenamefont {Fejer}}]{Nasr2008}%
  \BibitemOpen
  \bibfield  {author} {\bibinfo {author} {\bibfnamefont {M.}~\bibnamefont
  {Nasr}}, \bibinfo {author} {\bibfnamefont {S.}~\bibnamefont {Carrasco}},
  \bibinfo {author} {\bibfnamefont {B.}~\bibnamefont {Saleh}}, \bibinfo
  {author} {\bibfnamefont {A.}~\bibnamefont {Sergienko}}, \bibinfo {author}
  {\bibfnamefont {M.}~\bibnamefont {Teich}}, \bibinfo {author} {\bibfnamefont
  {J.}~\bibnamefont {Torres}}, \bibinfo {author} {\bibfnamefont
  {L.}~\bibnamefont {Torner}}, \bibinfo {author} {\bibfnamefont
  {D.}~\bibnamefont {Hum}}, \ and\ \bibinfo {author} {\bibfnamefont
  {M.}~\bibnamefont {Fejer}},\ }\href {\doibase 10.1103/PhysRevLett.100.183601}
  {\bibfield  {journal} {\bibinfo  {journal} {Physical Review Letters}\
  }\textbf {\bibinfo {volume} {100}},\ \bibinfo {pages} {183601} (\bibinfo
  {year} {2008})}\BibitemShut {NoStop}%
\bibitem [{\citenamefont {Dayan}\ \emph {et~al.}(2005)\citenamefont {Dayan},
  \citenamefont {Pe'er}, \citenamefont {Friesem},\ and\ \citenamefont
  {Silberberg}}]{Dayan2005}%
  \BibitemOpen
  \bibfield  {author} {\bibinfo {author} {\bibfnamefont {B.}~\bibnamefont
  {Dayan}}, \bibinfo {author} {\bibfnamefont {A.}~\bibnamefont {Pe'er}},
  \bibinfo {author} {\bibfnamefont {A.~A.}\ \bibnamefont {Friesem}}, \ and\
  \bibinfo {author} {\bibfnamefont {Y.}~\bibnamefont {Silberberg}},\ }\href
  {\doibase 10.1103/PhysRevLett.94.043602} {\bibfield  {journal} {\bibinfo
  {journal} {Physical Review Letters}\ }\textbf {\bibinfo {volume} {94}},\
  \bibinfo {pages} {043602} (\bibinfo {year} {2005})}\BibitemShut {NoStop}%
\bibitem [{\citenamefont {O'Donnell}\ and\ \citenamefont
  {U'Ren}(2009)}]{ODonnell2009}%
  \BibitemOpen
  \bibfield  {author} {\bibinfo {author} {\bibfnamefont {K.}~\bibnamefont
  {O'Donnell}}\ and\ \bibinfo {author} {\bibfnamefont {A.}~\bibnamefont
  {U'Ren}},\ }\href {\doibase 10.1103/PhysRevLett.103.123602} {\bibfield
  {journal} {\bibinfo  {journal} {Physical Review Letters}\ }\textbf {\bibinfo
  {volume} {103}},\ \bibinfo {pages} {123602} (\bibinfo {year}
  {2009})}\BibitemShut {NoStop}%
\bibitem [{\citenamefont {Z{\"{a}}h}\ \emph {et~al.}(2008)\citenamefont
  {Z{\"{a}}h}, \citenamefont {Halder},\ and\ \citenamefont {Feurer}}]{Zah2008}%
  \BibitemOpen
  \bibfield  {author} {\bibinfo {author} {\bibfnamefont {F.}~\bibnamefont
  {Z{\"{a}}h}}, \bibinfo {author} {\bibfnamefont {M.}~\bibnamefont {Halder}}, \
  and\ \bibinfo {author} {\bibfnamefont {T.}~\bibnamefont {Feurer}},\ }\href
  {http://www.ncbi.nlm.nih.gov/pubmed/18852751} {\bibfield  {journal} {\bibinfo
   {journal} {Optics express}\ }\textbf {\bibinfo {volume} {16}},\ \bibinfo
  {pages} {16452} (\bibinfo {year} {2008})}\BibitemShut {NoStop}%
\bibitem [{\citenamefont {Kuzucu}\ \emph {et~al.}(2008)\citenamefont {Kuzucu},
  \citenamefont {Wong}, \citenamefont {Kurimura},\ and\ \citenamefont
  {Tovstonog}}]{Kuzucu2008}%
  \BibitemOpen
  \bibfield  {author} {\bibinfo {author} {\bibfnamefont {O.}~\bibnamefont
  {Kuzucu}}, \bibinfo {author} {\bibfnamefont {F.}~\bibnamefont {Wong}},
  \bibinfo {author} {\bibfnamefont {S.}~\bibnamefont {Kurimura}}, \ and\
  \bibinfo {author} {\bibfnamefont {S.}~\bibnamefont {Tovstonog}},\ }\href
  {\doibase 10.1103/PhysRevLett.101.153602} {\bibfield  {journal} {\bibinfo
  {journal} {Physical Review Letters}\ }\textbf {\bibinfo {volume} {101}},\
  \bibinfo {pages} {153602} (\bibinfo {year} {2008})}\BibitemShut {NoStop}%
\bibitem [{\citenamefont {O'Donnell}(2011)}]{ODonnell2011}%
  \BibitemOpen
  \bibfield  {author} {\bibinfo {author} {\bibfnamefont {K.~a.}\ \bibnamefont
  {O'Donnell}},\ }\href {\doibase 10.1103/PhysRevLett.106.063601} {\bibfield
  {journal} {\bibinfo  {journal} {Physical Review Letters}\ }\textbf {\bibinfo
  {volume} {106}},\ \bibinfo {pages} {063601} (\bibinfo {year}
  {2011})}\BibitemShut {NoStop}%
\bibitem [{\citenamefont {Lukens}\ \emph {et~al.}(2013)\citenamefont {Lukens},
  \citenamefont {Dezfooliyan}, \citenamefont {Langrock}, \citenamefont {Fejer},
  \citenamefont {Leaird},\ and\ \citenamefont {Weiner}}]{Lukens2013}%
  \BibitemOpen
  \bibfield  {author} {\bibinfo {author} {\bibfnamefont {J.~M.}\ \bibnamefont
  {Lukens}}, \bibinfo {author} {\bibfnamefont {A.}~\bibnamefont {Dezfooliyan}},
  \bibinfo {author} {\bibfnamefont {C.}~\bibnamefont {Langrock}}, \bibinfo
  {author} {\bibfnamefont {M.~M.}\ \bibnamefont {Fejer}}, \bibinfo {author}
  {\bibfnamefont {D.~E.}\ \bibnamefont {Leaird}}, \ and\ \bibinfo {author}
  {\bibfnamefont {A.~M.}\ \bibnamefont {Weiner}},\ }\href {\doibase
  10.1103/PhysRevLett.111.193603} {\bibfield  {journal} {\bibinfo  {journal}
  {Physical Review Letters}\ }\textbf {\bibinfo {volume} {111}},\ \bibinfo
  {pages} {193603} (\bibinfo {year} {2013})}\BibitemShut {NoStop}%
\bibitem [{\citenamefont {Shaked}\ \emph {et~al.}(2014)\citenamefont {Shaked},
  \citenamefont {Pomerantz}, \citenamefont {Vered},\ and\ \citenamefont
  {Peʼer}}]{Shaked2014}%
  \BibitemOpen
  \bibfield  {author} {\bibinfo {author} {\bibfnamefont {Y.}~\bibnamefont
  {Shaked}}, \bibinfo {author} {\bibfnamefont {R.}~\bibnamefont {Pomerantz}},
  \bibinfo {author} {\bibfnamefont {R.~Z.}\ \bibnamefont {Vered}}, \ and\
  \bibinfo {author} {\bibfnamefont {A.}~\bibnamefont {Peʼer}},\ }\href
  {\doibase 10.1088/1367-2630/16/5/053012} {\bibfield  {journal} {\bibinfo
  {journal} {New Journal of Physics}\ }\textbf {\bibinfo {volume} {16}},\
  \bibinfo {pages} {053012} (\bibinfo {year} {2014})}\BibitemShut {NoStop}%
\bibitem [{\citenamefont {Harris}(2007)}]{Harris2007}%
  \BibitemOpen
  \bibfield  {author} {\bibinfo {author} {\bibfnamefont {S.~E.}\ \bibnamefont
  {Harris}},\ }\href {\doibase 10.1103/PhysRevLett.98.063602} {\bibfield
  {journal} {\bibinfo  {journal} {Physical Review Letters}\ }\textbf {\bibinfo
  {volume} {98}},\ \bibinfo {pages} {063602} (\bibinfo {year}
  {2007})}\BibitemShut {NoStop}%
\bibitem [{\citenamefont {Walls}\ and\ \citenamefont
  {Tindle}(1972)}]{Walls1972}%
  \BibitemOpen
  \bibfield  {author} {\bibinfo {author} {\bibfnamefont {D.~F.}\ \bibnamefont
  {Walls}}\ and\ \bibinfo {author} {\bibfnamefont {C.~T.}\ \bibnamefont
  {Tindle}},\ }\href {\doibase 10.1088/0305-4470/5/4/010} {\bibfield  {journal}
  {\bibinfo  {journal} {Journal of Physics A: General Physics}\ }\textbf
  {\bibinfo {volume} {5}},\ \bibinfo {pages} {534} (\bibinfo {year}
  {1972})}\BibitemShut {NoStop}%
\bibitem [{\citenamefont {Dowling}(2008)}]{Dowling2008}%
  \BibitemOpen
  \bibfield  {author} {\bibinfo {author} {\bibfnamefont {J.~P.}\ \bibnamefont
  {Dowling}},\ }\href {\doibase 10.1080/00107510802091298} {\bibfield
  {journal} {\bibinfo  {journal} {Contemporary Physics}\ }\textbf {\bibinfo
  {volume} {49}},\ \bibinfo {pages} {125} (\bibinfo {year} {2008})}\BibitemShut
  {NoStop}%
\bibitem [{\citenamefont {Brown}\ and\ \citenamefont
  {Twiss}(1957)}]{Brown1957}%
  \BibitemOpen
  \bibfield  {author} {\bibinfo {author} {\bibfnamefont {R.}~\bibnamefont
  {Brown}}\ and\ \bibinfo {author} {\bibfnamefont {R.}~\bibnamefont {Twiss}},\
  }\href {http://rspa.royalsocietypublishing.org/content/242/1230/300.short}
  {\bibfield  {journal} {\bibinfo  {journal} {Proceedings of the Royal Society
  of London. Series A. Mathematical and Physical Sciences}\ }\textbf {\bibinfo
  {volume} {242}},\ \bibinfo {pages} {300} (\bibinfo {year}
  {1957})}\BibitemShut {NoStop}%
\bibitem [{\citenamefont {Glauber}(1963)}]{Glauber1963}%
  \BibitemOpen
  \bibfield  {author} {\bibinfo {author} {\bibfnamefont {R.}~\bibnamefont
  {Glauber}},\ }\href {\doibase 10.1103/PhysRev.130.2529} {\bibfield  {journal}
  {\bibinfo  {journal} {Physical Review}\ }\textbf {\bibinfo {volume} {130}},\
  \bibinfo {pages} {2529} (\bibinfo {year} {1963})}\BibitemShut {NoStop}%
\bibitem [{\citenamefont {Loudon}\ and\ \citenamefont
  {Knight}(1987)}]{Loudon1987}%
  \BibitemOpen
  \bibfield  {author} {\bibinfo {author} {\bibfnamefont {R.}~\bibnamefont
  {Loudon}}\ and\ \bibinfo {author} {\bibfnamefont {P.}~\bibnamefont
  {Knight}},\ }\href {\doibase 10.1080/09500348714550721} {\bibfield  {journal}
  {\bibinfo  {journal} {Journal of Modern Optics}\ }\textbf {\bibinfo {volume}
  {34}},\ \bibinfo {pages} {709} (\bibinfo {year} {1987})}\BibitemShut
  {NoStop}%
\bibitem [{\citenamefont {Lefevre}(1980)}]{Lefevre1980}%
  \BibitemOpen
  \bibfield  {author} {\bibinfo {author} {\bibfnamefont {H.}~\bibnamefont
  {Lefevre}},\ }\href {\doibase 10.1049/el:19800552} {\bibfield  {journal}
  {\bibinfo  {journal} {Electronics Letters}\ }\textbf {\bibinfo {volume}
  {16}},\ \bibinfo {pages} {778} (\bibinfo {year} {1980})}\BibitemShut
  {NoStop}%
\bibitem [{\citenamefont {Lerch}\ and\ \citenamefont
  {Stefanov}(2014)}]{Lerch2014}%
  \BibitemOpen
  \bibfield  {author} {\bibinfo {author} {\bibfnamefont {S.}~\bibnamefont
  {Lerch}}\ and\ \bibinfo {author} {\bibfnamefont {A.}~\bibnamefont
  {Stefanov}},\ }\href {\doibase 10.1364/OL.39.005399} {\bibfield  {journal}
  {\bibinfo  {journal} {Optics Letters}\ }\textbf {\bibinfo {volume} {39}},\
  \bibinfo {pages} {5399} (\bibinfo {year} {2014})}\BibitemShut {NoStop}%
\bibitem [{\citenamefont {Bernhard}\ \emph {et~al.}(2013)\citenamefont
  {Bernhard}, \citenamefont {Bessire}, \citenamefont {Feurer},\ and\
  \citenamefont {Stefanov}}]{Bernhard2013}%
  \BibitemOpen
  \bibfield  {author} {\bibinfo {author} {\bibfnamefont {C.}~\bibnamefont
  {Bernhard}}, \bibinfo {author} {\bibfnamefont {B.}~\bibnamefont {Bessire}},
  \bibinfo {author} {\bibfnamefont {T.}~\bibnamefont {Feurer}}, \ and\ \bibinfo
  {author} {\bibfnamefont {A.}~\bibnamefont {Stefanov}},\ }\href {\doibase
  10.1103/PhysRevA.88.032322} {\bibfield  {journal} {\bibinfo  {journal}
  {Physical Review A}\ }\textbf {\bibinfo {volume} {88}},\ \bibinfo {pages}
  {032322} (\bibinfo {year} {2013})}\BibitemShut {NoStop}%
\end{thebibliography}
%	
%\printbibliography

\end{document}